\documentclass[aps,showpacs,nofootinbib,superscriptaddress,
preprintnumbers]{revtex4}
\usepackage{epsf,amsmath,amssymb,scalefnt}
\newcommand{\abbrev}{\scalefont{.9}}

\newcommand{\api}{\frac{\alpha_s}{\pi}}
\newcommand{\eqn}[1]{Eq.\,(\ref{#1})}
\newcommand{\fig}[1]{Fig.\,\ref{#1}}

\newcommand{\sct}[1]{Sect.\,\ref{#1}}
\newcommand{\dd}{{\rm d}}

\newcommand{\order}[1]{{\cal O}(#1)}
\newcommand{\lo}{{\abbrev LO}}
\newcommand{\nlo}{{\abbrev NLO}}
\newcommand{\nnlo}{{\abbrev NNLO}}
\newcommand{\qcd}{{\abbrev QCD}}

\newcommand{\reference}[1]{Ref.\,\cite{#1}}

\newcommand{\Msusy}[1]{M_{\tilde #1}}
\newcommand{\Mtop}{M_t}
\newcommand{\Mstop}{\Msusy{t}}
\newcommand{\Mgluino}{\Msusy{g}}
\newcommand{\lmut}{l_{\mu t}}
\newcommand{\sm}{{\abbrev SM}}
\newcommand{\mssm}{{\abbrev MSSM}}
\newcommand{\susy}{{\abbrev SUSY}}
\newcommand{\higgs}{\phi}
\newcommand{\lhc}{{\abbrev LHC}}
\newcommand{\muR}{\mu_{\rm R}}
\newcommand{\muF}{\mu_{\rm F}}
\newcommand{\msbar}{\overline{\mbox{\abbrev MS}}}

\begin{document}    

\preprint{CERN-TH/2003-195 --- DESY 03--109 --- hep-ph/0308210 --- August 2003}

\title{
Effects of {\abbrev SUSY-QCD} in hadronic Higgs production at 
next-to-next-to-leading order
\vskip.3cm
}
\author{Robert V. Harlander}
\affiliation{Theory Division, CERN, CH-1211 Geneva 23, Switzerland}
   \email{ robert.harlander@cern.ch }
\author{Matthias Steinhauser}
\affiliation{II. Institut f\"ur Theoretische Physik,
  \it Universit\"at Hamburg, D-22761 Hamburg, Germany}
  \email{ matthias.steinhauser@desy.de }

\begin{abstract}
  An estimate of the \nnlo{} supersymmetric {\abbrev QCD} effects for
  Higgs production at hadron colliders is given. Assuming an effective
  gluon--Higgs interaction, these corrections enter only in terms of
  process-independent, factorizable terms. We argue that the current
  knowledge of these terms up to \nlo{} is sufficient to derive the
  \nnlo{} hadronic cross section within the limitations of the standard
  theoretical uncertainties arising mainly from renormalization and
  factorization scale variations.  The \susy{} contributions are small
  with respect to the \qcd{} effects, which means that the \nnlo{}
  corrections to Higgs production are very similar in the \sm{} and the
  \mssm{}.
\end{abstract}

\maketitle

\section{Introduction}
Gluon fusion is the dominant production mechanism for Higgs bosons at
the Large Hadron Collider (\lhc{}) (for reviews, see
Refs.\,\cite{hunter,Spira:1997dg}).  A feature of the gluon fusion
process is that it is loop-mediated already at leading order. This makes
it particularly sensitive to non-standard particles and couplings as
they are predicted by extended theories.  A very popular extension of
the Standard Model (\sm{}) is the Minimal Supersymmetric Standard Model
(\mssm{}) (for a review, see \reference{Martin:1997ns}), with its five
physical Higgs bosons.

We will focus on a scenario where the ratio of the vacuum expectation
values of the two Higgs doublets is not too large, $\tan\beta\ll
M_t/M_b$, ($M_t$\,=\,top mass, $M_b$\,=\,bottom mass), so that the
bottom is much smaller than the top Yukawa coupling.  In this case, the
dominant effects on the gluon--Higgs coupling in the \mssm{} arise from
the top quark $t$ and its scalar supersymmetric (\susy{}) partner, the
top squark $\tilde t$. \susy{}-\qcd{} corrections are induced by virtual
gluons $g$ and their fermionic \susy{} partners, the gluinos $\tilde g$.
These effects have recently~\cite{Harlander:2003bb} (see also
\reference{Dawson:1996xz}) been evaluated at next-to-leading order
(\nlo{}) in the limit where $M_\higgs\ll \{\Mtop,\Mstop,\Mgluino\}$,
where $\phi$ denotes either of the two {\abbrev CP}-even Higgs bosons,
$h$ or $H$.  This limit is expected to work extremely well, if the
leading order (\lo{}) dependence on $\Mtop,\Mstop,\Mgluino$ is taken
into account exactly.  This can be inferred from the \nlo{} behavior in
the
\sm{}~\cite{Dawson:1990zj,Djouadi:1991tk,Graudenz:1992pv,Spira:1995rr}.

In the effective Lagrangian approach, the evaluation of the hadronic
Higgs cross section factorizes into the calculation of the effective
gluon--Higgs coupling, times the calculation of the actual process
$pp\to \higgs+X$ as mediated by the effective gluon--Higgs operator. For
a full next-to-next-to-leading order (\nnlo{}) result in this approach,
both factors need to be evaluated up to \nnlo{}. However, in the \sm{},
the \nnlo{} contribution of the effective coupling leads to a
numerically negligible contribution, and we will argue that this is true
also in the \mssm{}.  The \nnlo{} Higgs production cross section can
therefore be evaluated from the \nlo{} expression of the effective
coupling, as taken from \reference{Harlander:2003bb}, and the \nnlo{}
results for the process diagrams, which are identical to the \sm{}
case~\cite{1Harlander:2000mg,2Harlander:2002wh,3Anastasiou:2002yz,
  4Ravindran:2003um}.

\section{The approximation}\label{sec::app}

\subsection{Definition and Standard Model case}
We use the effective Lagrangian approach where the top quark and all
supersymmetric particles are considered heavy with respect to the Higgs
boson, see~\reference{Harlander:2003bb}.  In this case, the hadronic
cross section $\sigma_{hk} \equiv \sigma(hk\to \higgs+X)$ for Higgs
production can be written as
\begin{equation}
\begin{split}
  \sigma_{hk}(z) = \sigma_0\,C^2\,\Sigma_{hk}(z)\,,
\end{split}
\end{equation}
\begin{equation}
\begin{split}
\Sigma_{hk}(z) &= \sum_{i,j}
  \int_z^1\dd x_1 \int_{z/x_1}^1
  \dd x_2\,\,
  \varphi_{i/h}(x_1)\,
  \varphi_{j/k}(x_2)\,
  \hat\Sigma_{ij}\left(\frac{z}{x_1x_2}\right)\,,\qquad
  z\equiv \frac{M_\higgs^2}{s}\,,
  \label{eq::sigmahk}
\end{split}
\end{equation}
where $i$, $j$ denote any partons inside the hadrons $h$, $k$, and
$\varphi_{i/h}(x)$ are the parton densities; $M_\higgs$ is the Higgs
boson mass, and $s$ is the hadronic center-of-mass (c.m.) energy.

The coefficient function $C$, defined below, contains the remnant
dependence of the gluon--Higgs coupling on the heavy masses, and
$\sigma_0$ is defined such that the leading order dependence on these
masses of $\sigma_{hk}(z)$ is exact. Its exact form is irrelevant for
our argument and shall not be given here, owing to space limitations
(see, e.g.\ \reference{Harlander:2003bb}).

The partonic expression can be expanded in terms of $\alpha_s$:
\begin{equation}
\begin{split}
  \hat\Sigma_{ij}(x) = \hat\Sigma_{ij}^{(0)}(x) +
  \api\,\hat\Sigma_{ij}^{(1)}(x) +
  \left(\api\right)^2\hat\Sigma_{ij}^{(2)}(x) + \order{\alpha_s^3}\,,
\end{split}
\end{equation}
where $x\equiv M_\higgs^2/\hat s$, and $\hat s$ is the partonic
c.m.\ energy. Here and in what follows, $\alpha_s$
denotes the $\msbar$-renormalized strong coupling constant for five
active quark flavors, evaluated at the renormalization scale $\muR$.

For the coefficient function, we write
\begin{equation}
\begin{split}
  C(\alpha_s) = \api\,C^{(0)} \left[ 1 + \api\,\kappa_1 +
    \left(\api\right)^2\,\kappa_2 + \order{\alpha_s^3} \right]\,.
\label{eq::coef}
\end{split}
\end{equation}
In the \sm{}, the $\kappa_i$ are known for
$i=1,\ldots,3$~\cite{1Chetyrkin:1997iv,3Kramer:1996iq} ($\kappa_3$
contributes only at {\abbrev N}$^3${\abbrev LO}).  In the \mssm{}, $\kappa_1$
has been evaluated only recently~\cite{Harlander:2003bb}.

We now define
\begin{equation}
\begin{split}
\Sigma^{(n)}_{hk}(z) &= \sum_{i,j}
  \int_z^1\dd x_1 \int_{z/x_1}^1
  \dd x_2\,\,
  \varphi_{i/h}(x_1)\,
  \varphi_{j/k}(x_2)\,
  \hat\Sigma^{(n)}_{ij}\left(\frac{z}{x_1x_2}\right)\,, \qquad n\in
  \{0,1,2\}\,.
\end{split}
\end{equation}
For the $\Sigma^{(n)}_{hk}(z)$, $n=0,1,2$, we assume that the parton densities
$\varphi_{i/h}$ are evaluated at \nnlo{}.\footnote{We use the
  approximate \nnlo{} parton densities of \reference{Martin:2001es}.}
Thus, the \nnlo{} expression for the hadronic cross section can be written as
\begin{equation}
\begin{split}
  \sigma^{\rm NNLO} = \sigma_0\,\left(C^{(0)}\,\api\right)^2 \bigg[
  \Sigma^{(0)} &+ \api\,\left( \Sigma^{(1)} + 2\,\kappa_1\,\Sigma^{(0)}
  \right) \\& +\left(\api\right)^2\,\left( \Sigma^{(2)} +
    2\,\kappa_1\,\Sigma^{(1)} + (\kappa_1^2 + 2\,\kappa_2)\,\Sigma^{(0)}
  \right) \bigg]\,,
\label{eq::sigmannlo}
\end{split}
\end{equation}
where the indices $h,k\in\{p,\bar p\}$ have been dropped for
simplicity.
The basis of our estimate of the \nnlo{} terms in \susy{} will be that the
numerical effect of the term proportional to $\kappa_2$ in
\eqn{eq::sigmannlo} is negligible compared to the theoretical
uncertainty of the \nnlo{} prediction.

\begin{figure}
  \begin{center}
    \leavevmode
    \begin{tabular}{cc}
      \epsfxsize=18em
      \epsffile[110 265 465 560]{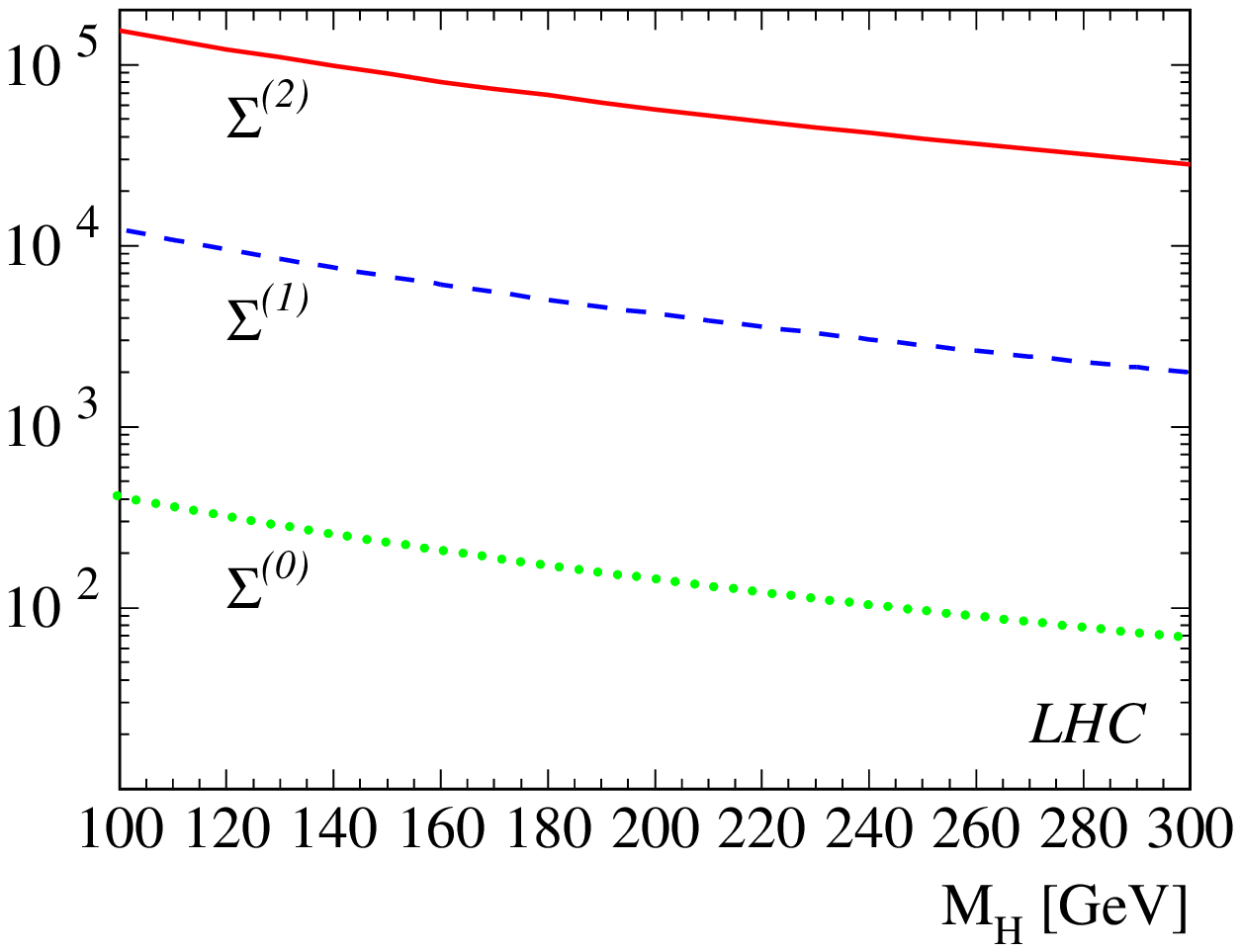} &
      \epsfxsize=18em
      \epsffile[110 265 465 560]{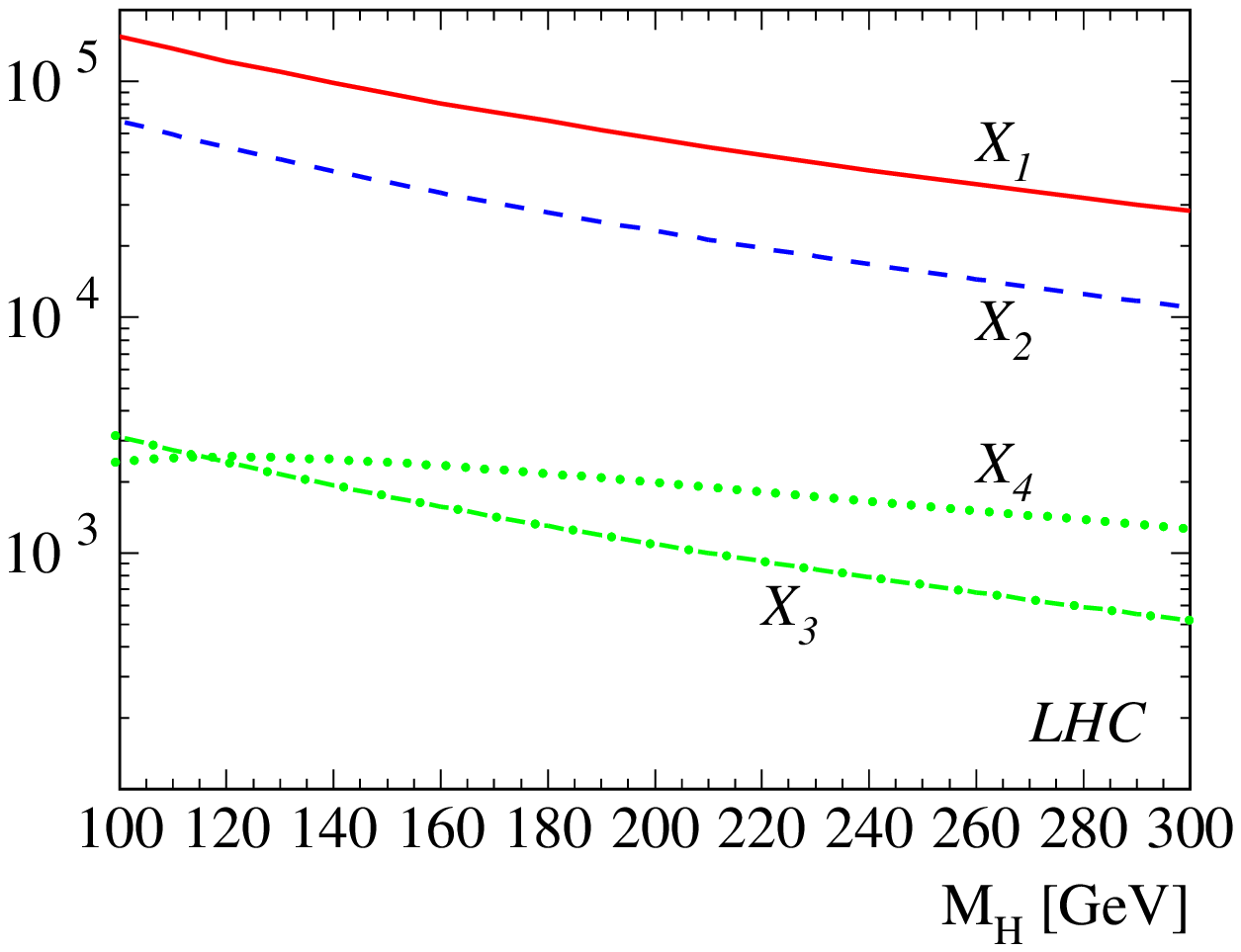} \\
      $(a)$ & $(b)$
    \end{tabular}
    \parbox{14.cm}{
      \caption[]{\label{fig::sig012}\sloppy
        Individual \nnlo{} contributions to the total hadronic Higgs
        production cross section. The notation is defined in
        Eqs.\,(\ref{eq::sigmannlo}) and (\ref{eq::x012def}).  The
        renormalization and factorization scale $\muR$ and $\muF$ are
        identified with the Higgs mass $M_\phi$.  }}
  \end{center}
\end{figure}

For that purpose, let us look at the relative magnitude of the
$\Sigma^{(n)}$ in the case of Higgs production at the \lhc{},
\fig{fig::sig012}\,$(a)$. We see that $\Sigma^{(0)}$ is more than two
orders of magnitude smaller than $\Sigma^{(2)}$, which suggests that the
effects from $\kappa_2$ can be neglected, if $\kappa_2$ is not too
large.  In order to get a feeling for the magnitude of $\kappa_2$, let
us look at the \sm{} case.  There we
have~\cite{1Chetyrkin:1997iv,3Kramer:1996iq}:
\begin{equation}
\begin{split}
C^{(0),\rm SM} &= -\frac{1}{3}\,,\qquad
\kappa_1^{\rm SM} = \frac{11}{4} = 2.75\,,\\
\kappa_2^{\rm SM} &= \frac{2777}{288} + \frac{19}{16}\,\lmut 
+ n_f\left( -\frac{67}{96} + \frac{1}{3}\,\lmut \right)
\stackrel{n_f=5}{\approx} 6.153 + 2.854\,\lmut\,,
\label{eq::cism}
\end{split}
\end{equation}
with $\lmut \equiv \ln(\muR^2/\Mtop^2)$, where $\muR$ is the
renormalization scale and $\Mtop$ is the on-shell top quark mass.

Using these numbers, we arrive at \fig{fig::sig012}\,$(b)$.  It shows
the relative size of the four terms that contribute to the cross section
in \eqn{eq::sigmannlo} at order $\alpha_s^4$:
\begin{equation}
\begin{split}
X_1 &= \Sigma^{(2)}\,,\qquad
X_2 = 2\,\kappa_1\,\Sigma^{(1)}\,,\qquad
X_3 = \kappa_1^2\,\Sigma^{(0)}\,,\qquad
X_4 = 2\,\kappa_2\,\Sigma^{(0)}\,.
\label{eq::x012def}
\end{split}
\end{equation}
As expected from the numerical value of $\kappa_2^{\rm SM}$, \eqn{eq::cism},
$X_4$ is indeed negligible with respect to $X_1$: it is down by a factor of
30. But another remarkable observation is that the term proportional to
$\Sigma^{(1)}$, i.e.\ $X_2$, amounts to around 30\% of the full
$\alpha_s^4$ contribution. For comparison: the $(2\,\kappa_1\,\Sigma^{(0)})$
term in \eqn{eq::sigmannlo} amounts to only 15\% of the complete
$\alpha_s^3$ contribution.

To summarize: In the \sm{}, the $\alpha_s^3$ term $\kappa_2$ to the
coefficient function of \eqn{eq::coef} gives a negligible contribution
to the \nnlo{} cross section. In fact, we checked that the difference
between the true\footnote{Within the effective theory approach.}  and
the approximate \nnlo{} cross section (i.e.\ with $\kappa_2=0$) is less
than $1\%$ at the \lhc{}. This is much smaller than the theoretical uncertainty
of around 15\%, as estimated by the variation of the factorization and the
renormalization scale at \nnlo{}.  On the other hand, the knowledge of
$\kappa_1$ is, relatively speaking, numerically more important for the
\nnlo{} than for the \nlo{} contribution to the cross section, for which
it was originally evaluated~\cite{Harlander:2003bb}.

\subsection{Minimal Supersymmetric Standard Model}
In the \mssm{}, we can parameterize the \nlo{} corrections to the effective
Lagrangian as
\begin{equation}
\begin{split}
  \kappa_1^{\rm SUSY} = \kappa_1^{\rm SM} + \delta\kappa_1 = \frac{11}{4} +
  \delta\kappa_1\,.
\end{split}
\end{equation}
In addition, the tree-level normalization of \eqn{eq::coef}, $C^{(0)}$,
changes, of course, but this is irrelevant for our discussion.
$\delta\kappa_1$ was recently computed in
\reference{Harlander:2003bb}\footnote{In the notation of
  \reference{Harlander:2003bb}, it is $c^{\rm SUSY} = \delta\kappa_1 +
  \order{\alpha_s}$.} 
and was shown to be negative, with
\begin{equation}
\begin{split}
|\delta\kappa_1| \lesssim 1
\end{split}
\end{equation}
for relevant values of the \susy{} parameters (recall that we restrict
ourselves to $\tan\beta \ll \Mtop{}/M_b$).  It is thus reasonable to
assume that also the value of $\kappa_2$ in {\abbrev SUSY-QCD} will be
of the same order of magnitude as in the \sm{} (or smaller).  Combining
this assumption with the discussion of \fig{fig::sig012} (see above)
leads us to the conclusion that the \nnlo{} cross section for hadronic
Higgs production in supersymmetry should be approximated well by setting
$\kappa_2\approx\kappa_2^{\rm SM}$.

\section{Results}
As in \reference{Harlander:2003bb}, we will neglect squark mixing
and set the bottom Yukawa coupling to zero for simplicity.
More detailed phenomenological studies have to be
deferred to a forthcoming publication.

\begin{figure}
  \begin{center}
    \leavevmode
    \begin{tabular}{c}
      \epsfxsize=20em
      \epsffile[110 265 465 560]{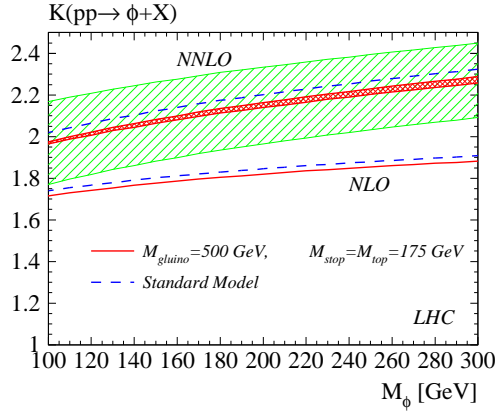}
    \end{tabular}
    \parbox{14.cm}{
      \caption[]{\label{fig::kss14nnlo}\sloppy
        K-factor in the Standard Model (dashed blue) and the \mssm{}
        (solid red) for the indicated set of parameters. The narrow
        (red) band in the \nnlo{} \mssm{} curve corresponds to varying
        $\kappa_2$ between zero and $2\kappa_2^{\rm SM}$. The
        renormalization and factorization scales ($\muR$, $\muF$) have
        been identified with the Higgs mass in these curves. The
        diagonally shaded band (green) corresponds to the variation of
        $\muR$ between $2M_\higgs$ and $M_\phi/2$ in the \nnlo{} result
        with $\kappa_2=\kappa_2^{\rm SM}$ ($\muF=M_\phi$).}}
  \end{center}
\end{figure}

\fig{fig::kss14nnlo} shows the \nlo{} and the \nnlo{} K-factor, $K_{\rm
  X} \equiv \sigma^{\rm X}/\sigma^{\rm LO}$ (${\rm X} = {\rm NLO}, {\rm
  NNLO}$) in the \sm{} case (dashed), and in the \mssm{}, for $\Msusy{t}
= \Mtop = 175$\,GeV, and $\Mgluino = 500$\,GeV;
$\sigma^{\rm LO}$, $\sigma^{\rm NLO}$, and $\sigma^{\rm NNLO}$ are
evaluated with \lo{}, \nlo{}, and \nnlo{} parton densities and
$\alpha_s$ evolution.

The \nnlo{} result in the \mssm{} is given by the narrow (red) band,
arising from the variation of $\kappa_2$ between zero and
$2\,\kappa_2^{\rm SM}$ (see \eqn{eq::cism}).  This should serve as an
estimate of the theoretical uncertainty induced by the approximation
introduced in \sct{sec::app}.  Within our approximations, the K-factor
in \fig{fig::kss14nnlo} is valid for both {\abbrev CP}-even Higgs bosons
of the \mssm{}, since the Yukawa coupling cancels in the ratio of the
\lo{} to the higher order results.

As at \nlo{}, the \susy{} effects are small with respect to the \qcd{}
effects at \nnlo{}, so that the total K-factor in the \mssm{} is
very similar to its \sm{} value. The theoretical uncertainties due to
variation of the renormalization and factorization scales are almost
identical to the \sm{} case, since the only source of additional scale
dependence in the \mssm{} arises from $\kappa_2$. For
$\kappa_2=\kappa_2^{\rm SM}$, the $\muR$ dependence of the \nnlo{}
prediction is indicated in \fig{fig::kss14nnlo} as the diagonally shaded
band; the $\muF$ dependence is much smaller.

\section{Conclusions}
We have discussed an approximation for the \nnlo{} contributions to
supersymmetric Higgs production in gluon fusion. It was argued that the
corresponding \nnlo{} estimate should be accurate up to a few per cent,
which is much smaller than the theoretical uncertainty induced by the
residual renormalization and factorization scale dependence, as well as
the anticipated experimental accuracy.  For a given set of \susy{}
parameters, and within the restrictions on these parameters as discussed
in the main text, the production cross section for a Higgs boson in the
\mssm{} is thus known to a precision similar to that in the \sm{}.  More
detailed studies of the \mssm{} parameter space are clearly desirable
and will be presented elsewhere.

{\it Acknowledgements.}
We would like to thank T.~Plehn for carefully reading the manuscript.

\def\app#1#2#3{{\it Act.~Phys.~Pol.~}\jref{\bf B #1}{#2}{#3}}
\def\apa#1#2#3{{\it Act.~Phys.~Austr.~}\jref{\bf#1}{#2}{#3}}
\def\annphys#1#2#3{{\it Ann.~Phys.~}\jref{\bf #1}{#2}{#3}}
\def\cmp#1#2#3{{\it Comm.~Math.~Phys.~}\jref{\bf #1}{#2}{#3}}
\def\cpc#1#2#3{{\it Comp.~Phys.~Commun.~}\jref{\bf #1}{#2}{#3}}
\def\epjc#1#2#3{{\it Eur.\ Phys.\ J.\ }\jref{\bf C #1}{#2}{#3}}
\def\fortp#1#2#3{{\it Fortschr.~Phys.~}\jref{\bf#1}{#2}{#3}}
\def\ijmpc#1#2#3{{\it Int.~J.~Mod.~Phys.~}\jref{\bf C #1}{#2}{#3}}
\def\ijmpa#1#2#3{{\it Int.~J.~Mod.~Phys.~}\jref{\bf A #1}{#2}{#3}}
\def\jcp#1#2#3{{\it J.~Comp.~Phys.~}\jref{\bf #1}{#2}{#3}}
\def\jetp#1#2#3{{\it JETP~Lett.~}\jref{\bf #1}{#2}{#3}}
\def\jhep#1#2#3{{\small\it JHEP~}\jref{\bf #1}{#2}{#3}}
\def\mpl#1#2#3{{\it Mod.~Phys.~Lett.~}\jref{\bf A #1}{#2}{#3}}
\def\nima#1#2#3{{\it Nucl.~Inst.~Meth.~}\jref{\bf A #1}{#2}{#3}}
\def\npb#1#2#3{{\it Nucl.~Phys.~}\jref{\bf B #1}{#2}{#3}}
\def\nca#1#2#3{{\it Nuovo~Cim.~}\jref{\bf #1A}{#2}{#3}}
\def\plb#1#2#3{{\it Phys.~Lett.~}\jref{\bf B #1}{#2}{#3}}
\def\prc#1#2#3{{\it Phys.~Reports }\jref{\bf #1}{#2}{#3}}
\def\prd#1#2#3{{\it Phys.~Rev.~}\jref{\bf D #1}{#2}{#3}}
\def\pR#1#2#3{{\it Phys.~Rev.~}\jref{\bf #1}{#2}{#3}}
\def\prl#1#2#3{{\it Phys.~Rev.~Lett.~}\jref{\bf #1}{#2}{#3}}
\def\pr#1#2#3{{\it Phys.~Reports }\jref{\bf #1}{#2}{#3}}
\def\ptp#1#2#3{{\it Prog.~Theor.~Phys.~}\jref{\bf #1}{#2}{#3}}
\def\ppnp#1#2#3{{\it Prog.~Part.~Nucl.~Phys.~}\jref{\bf #1}{#2}{#3}}
\def\rmp#1#2#3{{\it Rev.~Mod.~Phys.~}\jref{\bf #1}{#2}{#3}}
\def\sovnp#1#2#3{{\it Sov.~J.~Nucl.~Phys.~}\jref{\bf #1}{#2}{#3}}
\def\sovus#1#2#3{{\it Sov.~Phys.~Usp.~}\jref{\bf #1}{#2}{#3}}
\def\tmf#1#2#3{{\it Teor.~Mat.~Fiz.~}\jref{\bf #1}{#2}{#3}}
\def\tmp#1#2#3{{\it Theor.~Math.~Phys.~}\jref{\bf #1}{#2}{#3}}
\def\yadfiz#1#2#3{{\it Yad.~Fiz.~}\jref{\bf #1}{#2}{#3}}
\def\zpc#1#2#3{{\it Z.~Phys.~}\jref{\bf C #1}{#2}{#3}}
\def\ibid#1#2#3{{ibid.~}\jref{\bf #1}{#2}{#3}}

\newcommand{\arxiv}[1]{{\tt #1}}
\newcommand{\jref}[3]{{\bf #1}, #3 (#2)}
\newcommand{\bibentry}[4]{#1, #3.}
\newcommand{\bibbook}[3]{#1, {\it #2}, #3.}

\end{document}